\documentstyle[12pt]{article}
\newcommand{\e}{\; {\rm e} }
\newcommand{\sgn}{\; {\rm sgn} }
\newcommand{\tr}{\; {\rm tr} }
\newcommand{\ch}{\; {\rm ch} }
\newcommand{\th}{\; {\rm th} }
\newcommand{\sh}{\; {\rm sh} }

\newcommand{\be}{\begin{eqnarray} }
\newcommand{\ee}{\end{eqnarray} }

\begin{document}
\begin{center}
\Large{\bf Chiral And Parity Anomalies At Finite Temperature And Density\\}
\vskip 5mm
\large{ A. N. Sissakian\\}
{\it Bogolubov Theoretical Laboratory,
Joint Institute for Nuclear Research,
Dubna, Moscow region 141980, Russia\\}
\vskip 5mm
\large{O. Yu. Shevchenko\footnote{shevch@nusun.jinr.ru}
and S. B. Solganik\footnote{solganik@thsun1.jinr.ru}\\}
{\it Laboratory of Nuclear Problems,
Joint Institute for Nuclear Research,
Dubna, Moscow region 141980, Russia\\}
\end{center}
\vskip 5mm

\begin{abstract}
Two closely related topological phenomenons are studied
at finite density and temperature.
These are chiral anomaly and Chern--Simons term.
By using different methods
it is shown  that $\mu^2 = m^2$ is the
crucial point for Chern--Simons at zero temperature.
So when $\mu^2 < m^2$  $\mu$--influence disappears and
we get the usual Chern-Simons term.
On the other hand when $\mu^2 > m^2$ the Chern-Simons term
vanishes because of non--zero density of background fermions.
It is occurs that the chiral anomaly doesn't depend on
density and temperature.
The connection between parity anomalous Chern-Simons and
chiral anomaly is generalized on finite density.
These results hold in any  dimension as in abelian,
so as in nonabelian cases.
\end{abstract}
\vskip 5mm

\section{Introduction}

Topological objects in modern physics play a great role.
In particular, here we are interested in Chern-Pontriagin and
Chern-Simons (CS) secondary characteristic classes.
That corresponds to chiral anomaly in even dimensions and
to CS (parity anomaly) in odd dimensions.
Both phenomenons are very important in quantum physics.
So, chiral anomalies in quantum field theory have  direct
applications to the decay of $\pi_{0}$ into two photons
($\pi_{0}\rightarrow\gamma\gamma$), in the understanding and solution
of the U(1) problem and so on.
On the other hand, there are many
effects caused by CS secondary
characteristic class. These
are, for example, gauge particles mass appearance  in quantum
field theory, applications to condense matter physics  such as
the fractional quantum Hall effect and high $T_{c}$ superconductivity,
possibility of free of metric tensor theory construction
etc.

It must be emphasized that these two phenomenons are closely
related. As it was shown (at zero density) in \cite{ni,ni1}
the trace identities
connect even dimensional anomaly with the odd dimensional
CS.
The main goal of this paper is to explore these  anomalous
objects  at finite density and temperature.

It was shown \cite{niemi,redl,witten} in a conventional zero density
and temperature gauge
theory that the CS term is generated in the
Eulier--Heisenberg effective action
by quantum corrections.
Since  the chemical potential term $\mu\bar\psi\gamma^{0}\psi$ is
odd under charge conjugation we can expect that it would
contribute to $P$ and $CP$ nonconserving quantity ---  CS term.
As we will see, this expectation is completely justified.
The zero density approach usually is a good quantum field  approximation
when the chemical potential is small as compared with
characteristic energy scale  of physical processes.
Nevertheless, for investigation of topological effects
it is not the case.
As we will see below, even a small density could lead to
principal effects.

In the excellent paper by Niemi \cite{ni} it was emphasized that the
charge density
at $\mu \not = 0$ becomes nontopological object, i.e contains as topological
part so as nontopological one.
The charge density at $\mu \not = 0$ (nontopological, neither parity odd
nor parity even object)\footnote{For abbreviation,  speaking about parity
invariance properties of local objects, we will
keep in mind  symmetries of the corresponding action parts.}
in $QED_{3}$ at finite density
was calculated and exploited in \cite{tolpa}. It must be
emphasized that in \cite{tolpa} charge density
(calculated in the constant pure magnetic field)
contains
as well parity odd part corresponding to CS term,
so as parity even part, which can't be covariantized
and don't contribute to the mass of the gauge field.
Here we are interested in  finite density and temperature
influence
on covariant parity odd form in action leading to the
gauge field mass generation --- CS
topological term. Deep insight on this phenomena at small densities
was done in \cite{ni,ni1}.
The result for CS term coefficient in $QED_{3}$  is
$\left[ \th \frac{1}{2}\beta(m-\mu)+\th \frac{1}{2}\beta(m+\mu)\right]$
(see \cite{ni1}, formulas (10.18) ).
However, to get this result it was heuristicaly supposed
that at small densities index theorem could still be used and
only odd in energy part of spectral density is responsible for
parity nonconserving effect.
Because of this in \cite{ni1} it had been stressed
that the result holds only for small $\mu$. However,
as we'll see below this result holds for any values of
chemical potential.
Thus, to obtain trustful result at any values of $\mu$ one
have to use transparent and free of any restrictions on $\mu$
procedure,
which would allow to perform calculations
with arbitrary nonabelian background gauge fields.

It was shown at zero chemical potential in \cite{ni,ni1,niemi}
that CS term in odd dimensions
is connected with chiral anomaly in even dimensions by
trace identities.
As we'll see below it is possible to generalize
a trace identity on nonzero density case.
The trace identity connects chiral anomaly with CS term
which has $\mu$ and $T$ dependent coefficient.
Despite chemical potential and temperature
give rise to a coefficient in front of CS term
they doesn't influence on chiral anomaly.
Indeed, anomaly is  short distance phenomenon which should not
be effected by medium $\mu$ and $T$ effects, or more quantitatively,
so as  the anomaly has ultraviolet nature, temperature and chemical
potential should not give any ultraviolet effect since
distribution functions decrease exponentially with energy
in the ultraviolet limit.

This paper is organized as follows. In sect.2
the independence  of chiral anomaly on temperature and
background fermion density is discussed.
It is shown in 2-dimensional Schwinger model
that chiral anomaly isn't influenced not only by chemical
potential $\mu$,
but also by Lagrange multiplier $\kappa$ at
conservation of chiral charge constraint.
Besides, we consider CS term appearance at finite density
in even dimensional theories. In sect.3 we obtain CS term
in 3-dimensional theory at finite density and temperature
by use of a few different methods. In sect.4 we evaluate CS term
coefficient in 5-dimensional theory and generalize this result on
arbitrary nonabelian odd-dimensional theory.
In sect.5 we generalize trace
identity on finite density on the basis of the previous calculations.
Section 6 is devoted to concluding remarks.

\section{Chiral anomaly  and Chern-Simons term in even dimensions}

As it is well-known, chemical potential can be introduced in a theory
as Lagrange multiplier at corresponding conservation laws.
In nonrelativistic physics this is  conservation of full number of
particles. In relativistic quantum field theory these are the conserving
charges.
The ground state energy
can be obtained by use of variational principle
\be
\langle \psi^{*} \hat{H} \psi \rangle =min
\ee
under charge conservation constraint for relativistic
equilibrium system
\be
\label{ch1}
\langle \psi^{*} \hat{Q} \psi \rangle ={\it const},
\ee
where $\hat{H}$ and $\hat{Q}$ are
hamiltonian and charge operators. Instead, we can use method of
undetermined Lagrange multipliers and seek absolute minimum of
expression
\be
\langle \psi^{*} (\hat{H}-\mu\hat{Q}) \psi \rangle ,
\ee
where $\mu$ is Lagrange multiplier.
Since $\hat{Q}$ commute with the hamiltonian
$\langle\hat{J}_{0}\rangle$ is conserved.

On the other hand, we can impose
another constraint, which implies chiral charge conservation
\be
\langle \psi^{*} \hat{Q}_{5} \psi \rangle ={\it const},
\ee
i.e. in Lagrange approach we have
\be
\langle \psi^{*} (\hat{H}-\kappa\hat{Q_{5}}) \psi \rangle = {\rm min},
\ee
where $\kappa$ arises as Lagrange multiplier at
$\langle\hat{J}_{0}^{5}\rangle = const$ constraint.
Thus, $\mu$ corresponds to nonvanishing fermion density
(number of particles minus number of antiparticles)
in background. Meanwhile, $\kappa$ is responsible for conserving
asymmetry in numbers of left and right handed background fermions.

It must be emphasized that the
formal addition of a chemical potential in the theory
looks like a simple gauge transformation
with the gauge function $\mu t$. However, it doesn't only shift the time
component
of a vector potential but also gives corresponding prescription for
handling Green's function poles.
The correct introduction of a chemical potential redefines
the ground state (Fermi energy),
which leads to a new spinor propagator with the correct
$\epsilon$-prescription for poles.
So, for the free spinor propagator we have
(see, for example, \cite{shur,chod})
\begin{eqnarray}
\label{sh}
G(p;\mu)=
\frac{\tilde{\not\! p}+m}
{(\tilde{p_{0}}+i\epsilon\sgn p_0 )^2-\vec{p}^2-m^2 },
\end{eqnarray}
where $\tilde{p}=(p_0 + \mu,\vec{p})$. Thus, when $\mu =0$ one at once
gets the usual $\epsilon$-prescription  because of the positivity of
$p_0\sgn p_0$. In Euclidian metric one has
\begin{eqnarray}
\label{euc}
G(p;\mu)=
\frac{\tilde{\not\! p}+m}
{\tilde{p_{0}}^2+\vec{p}^2+m^2 },
\end{eqnarray}
where $\tilde{p}=(p_0 + i\mu,\vec{p})$.
In the presence of a  background Yang--Mills field we consequently
have for the Green function operator (in Minkovsky space)
\begin{eqnarray}
\label{gre}
\hat{G}=(\gamma \tilde{\pi}-m)\frac{1}{(\gamma\tilde{\pi})^2-m^2 + i \epsilon
(p_{0}+\mu)\sgn (p_{0})} ,
\end{eqnarray}
where $\tilde{\pi_{\nu}}=\pi_{\nu}+\mu\delta_{\nu 0}$ ,
$\pi_{\nu}=p_{\nu}-gA_{\nu}(x)$.

Now we'll consider chiral anomaly.
It was shown in \cite{chirm}, that chiral anomaly doesn't depend on
$\mu$ and $T$.
In \cite{chirm} the direct calculations in 4-dimensional gauge theory
were performed by use of
imaginary and real time formalism, by using Fujikawa method 
and perturbation theory. These calculations are rather
cumbersome. To clear understand the nature of anomaly $\mu$-independence
($T$-independence will be discussed later)
we'll consider here the simplest case -- 2-dimensional QED and rederive
result of \cite{chirm} by use of Schwinger nonperturbative
method \cite{shw}.
So, one can write
\be
J^{\mu}=-ig\tr\left[ \gamma^{\mu} G(x,x^{'})\exp\left(
-ig\int_{x^{'}}^{x}d\xi^{\mu}A_{\mu}(\xi)\right)\right]_{x^{'}
\rightarrow x},
\ee
where $G(x,x^{'})$ -- propagator satisfying following equation
\be
\gamma^{\mu}\left(\partial_{\mu}^{x}-igA_{\mu}(x)\right)G(x,x^{'})
=\delta (x-x^{'}).
\ee
Following Shcwinger we use anzats
\be
G(x,x^{'})=G^{0}(x,x^{'})\exp\left[ ig(\phi (x)-\phi(x^{'}))\right],
\ee
where $G^{0}(x,x^{'})$ -- free propagator
$$\gamma^{\mu}\partial_{\mu}^{x}G^{0}(x,x^{'})=\delta (x-x^{'}).$$
Thus, for $\phi$ we can write $\gamma^{\mu}\partial_{\mu}\phi=
\gamma^{\mu}A_{\mu}$. From (\ref{sh}) we have
\be
G^{0}(x,x^{'})&=&\int\frac{d^{2}p}{(2\pi)^{2}} \e^{ip(x-x^{'})}
\frac{\not\! p}{p^{2}+i\varepsilon (p_{0}+\mu)\sgn p_{0}}
=-i\not\! \partial\Biggl[ \int\frac{d^{2}p}{(2\pi)^{2}} \e^{ip(x-x^{'})}
\frac{1}{p^{2}+i\varepsilon } -\nonumber\\
&-&2\int_{-\infty}^{+\infty}\frac{dp_{1}}{2\pi}
\int_{-\infty}^{+\infty}\frac{dp_{0}}{2\pi}
\theta (-\tilde{p}_{0}\sgn p_{0})\e^{ip(x-x^{'})}\Im m
\frac{1}{p^{2}+i\varepsilon }\Biggr].
\ee
So, beside the usual zero density part $\mu$--dependent one appears.
Further, we have to take off regularization in the current by use of 
symmetrical limit $x\rightarrow x^{'}$. After some simple calculations it is 
clearly seen that all $\mu$--dependent terms after taking off the limit 
disappear. Thus, contribution to the current arises from the 
$\mu$-independent part only. So
\be
J^{\mu} &=& i\frac{g^2}{2\pi}\left(\delta^{\mu\nu} -
\frac{\partial^{\mu}\partial^{\nu}}{\partial^{2}}\right)A_{\nu},\nonumber\\
J^{\mu}_{5} &=& i\frac{g^2}{2\pi}\left(\varepsilon^{\mu\nu} -
\varepsilon^{\mu\alpha}
\frac{\partial^{\alpha}\partial^{\nu}}{\partial^{2}}\right)A_{\nu}
\ee
and we get the usual anomaly in the chiral current
\be
\partial_{\mu}J^{\mu}=0 \;\; , \;\;\;\;\;
\partial_{\mu}J_{5}^{\mu}=i\frac{g^{2}}{2\pi}\varepsilon^{\mu\nu}
\partial_{\mu}A_{\nu}=i\frac{g^{2}}{4\pi}\*^{*}F.
\ee

Let's now consider $\kappa$ influence on the chiral anomaly.
Since, as we've seen above, $\kappa$ is directly connected to
chiral charge  it would be natural to expect some $\kappa$
effect on chiral anomaly.
However, the rather amazing situation occurs.
The demand of chiral charge conservation
(instead of the usual charge conservation)
on the quantum level doesn't influence on chiral anomaly.
Really, in 2-dimensions  introduction of Lagrange multiplier $\kappa$
at the chiral charge conservation gives the term
$\kappa\bar\psi \gamma^{5}\gamma^{0}\psi=
\kappa\bar\psi \gamma^{1}\psi$ in lagrangian.
So,  $\kappa$ effects in the same way as $\mu$, i.e.
$\kappa$ doesn't influence on the chiral anomaly
(it is also seen from direct calculations which are similar
to presented above for the case with $\mu$).
That could be explained due to ultraviolet nature
of the chiral anomaly, while $\kappa$ ($\mu$) doesn't introduce
new divergences in the theory.

From the above calculations  it is clearly seen the principle difference
of chiral anomaly and CS. The ultraviolet regulator -- $P$-exponent
gives rise to the anomaly, but (as we'll see below) doesn't
influence on CS. Thus, it is natural, that the anomaly doesn't
depend on $\mu$, $\kappa$ and $T$ because
it has ultraviolet regularization
origin, while
neither density nor temperature does influence on
ultraviolet behavior of the theory. The general and clear proof
of axial anomaly temperature independence will be presented in
sect.5 on the basis of the trace identities.


We now consider CS  in even dimensional theory.
From the definition  one has
\be
\frac{\partial I_{eff}}{\partial\kappa} =
\int d^{D}x\langle J_{5}^{0}\rangle.
\ee
Since axial anomaly doesn't depend on $\kappa$,
effective action  contains the term proportional to anomalous
$Q_{5}$ charge with $\kappa$ as a coefficient. The same is for a
chiral theory.  There,  effective action contains the term
proportional to anomalous
$Q$ charge with $\mu$ as coefficient, see for example \cite{redl85,rub,rut}.
So, we have
\be
\label{z}
\Delta I_{eff} = -\kappa\int dx_{0} W[A]
\ee
in conventional gauge theory and
\be
\label{z1}
\Delta I_{eff}^{\rm chiral} =-\mu\int dx_{0} W[A]
\ee
in the chiral theory. Here $W[A]$ -- CS term.
Thus we get CS with Lagrange multiplier as a coefficient.

It is well-known that at nonzero temperature
in $\beta\rightarrow 0$ limit
the dimensional reduction effect occurs.
So, extra $t$-dependence of CS term in (\ref{z}) disappears
and CS can be treated as
a mass term in 3-dimensional theory with $i\kappa /T$ coefficient
(the same for chiral theory with $\mu$ see \cite{redl85}).
For anomalous parts of effective action we have
\be
\label{m3}
\Delta I_{eff} = -i\kappa\beta W[A] \;\;, \;\;\;\;\;\;
\Delta I_{eff}^{\rm chiral} =-i\mu\beta  W[A]
\ee
in conventional and chiral gauge theories correspondingly.
The only problem arises in a treating CS as
a mass term is that  the coefficient is imaginary,
see discussions on the theme in \cite{redl85,rut}.
One can notice that results (\ref{z}),(\ref{z1}) and (\ref{m3})
hold in arbitrary even dimension.
Let us stress that we don't need any complicated calculations
to obtain (\ref{z}--\ref{m3}). The only thing we need is
the knowledge of  chiral anomaly independence on $\mu$, $\kappa$
and $\beta$.

\section{CS in 3-dimensional theory}
\subsection{Constant magnetic field}

Let's first consider a (2+1) dimensional  abelian theory.
Here we'll use  constant magnetic background.
We'll evaluate
fermion density by performing
the direct summation over Landau levels.
As a starting point, we'll use the formula for fermion number at
finite density and temperature \cite{ni}
\be
\label{num}
N&=&- \frac{1}{2}\sum_{n}\th (\frac{1}{2}\beta\lambda_{n}) +
\sum_{n}\left[ \frac{\theta (\lambda_{n})}{\exp
(-\beta(\mu -\lambda_{n}))+1} -
\frac{\theta (-\lambda_{n})}{\exp (-\beta(\lambda_{n} -\mu ))+1}\right]
=\nonumber\\&=&
\frac{1}{2}\sum_{n}\th \frac{1}{2}\beta(\mu -\lambda_{n})
\stackrel{\beta\rightarrow\infty}{\longrightarrow}
\frac{1}{2}\sum_{n}\sgn (\mu -\lambda_{n}) .
\ee
Landau levels in the constant magnetic field have the form \cite{lan}
\be
\lambda_{0} = -m\sgn (eB)~~~,~~~~~~\lambda_{n}=\pm
\sqrt{2n|eB|+m^2}
\ee
where n=1,2, ...
It is also necessary to take into account in (\ref{num}) the degeneracy
of  Landau levels.
Namely, the number of degenerate states for each Landau level
is $|eB|/2\pi$ per unit  area.
Even now we can see that only zero modes
(because of $\sgn (eB)$)
could contribute to the parity odd quantity.
So, for zero temperature, by using the identity
\be
\sgn (a - b) +\sgn (a + b)=2\sgn (a) \theta (|a| -|b|),\nonumber
\ee
one gets for zero modes
\be
\frac{|eB|}{4\pi}\sgn \left(\mu + m\sgn (eB)\right)=
\frac{|eB|}{4\pi}\sgn (\mu) \theta (|\mu | -|m|) +
\frac{|eB|}{4\pi}\sgn (eB)\sgn (m) \theta (|m | -|\mu|),
\ee
and for nonzero modes
\be
\frac{1}{2}\frac{|eB|}{2\pi}\sum_{n=1}^{\infty}\sgn (\mu - \sqrt{2n|eB|+m^2})
+\sgn (\mu + \sqrt{2n|eB|+m^2})=\nonumber\\
=\frac{|eB|}{2\pi}\sgn (\mu) \sum_{n=1}^{\infty}\theta (|\mu | -
\sqrt{2n|eB|+m^2}).
\ee
Combining contributions of all modes we get for fermion  density
\be
\label{rho}
\rho&=&\frac{|eB|}{2\pi}\sgn (\mu) \sum_{n=1}^{\infty}
\theta\left(|\mu | - \sqrt{2n|eB|+m^2}\right) +
\frac{1}{2}\frac{|eB|}{2\pi}\sgn (\mu) \theta (|\mu | -|m|) +\nonumber\\
&+&\frac{1}{2}\frac{eB}{2\pi}\sgn (m) \theta (|m | -|\mu|)=\nonumber\\
&=&\frac{|eB|}{2\pi}\sgn (\mu)\left( {\rm Int}\left[\frac{\mu^2- m^2}{2 |eB|}
\right]+
\frac{1}{2}\right)\theta (|\mu|-|m|) + \frac{eB}{4\pi}\sgn (m)
\theta (|m | -|\mu|).
\ee
Here we see that zero modes contribute as to parity odd so as to parity
even part, while  nonzero modes contribute to the parity even part only
(note that under parity transformation $B\rightarrow -B$).
Thus, fermion  density  contains as parity odd part leading
to CS term in action after covariantization,
so as parity even part.
It is straightforward to generalize the calculations on
finite temperature case.
Substituting zero modes into (\ref{num}) one gets
\be
N_{0}&=&\frac{|eB|}{2\pi}\frac{1}{2}\th\left[ \frac{1}{2}\beta
\left( \mu + m\sgn (eB) \right)  \right] =\nonumber\\&=&
\frac{|eB|}{4\pi}\left[ \frac{\sh (\beta\mu)}{\ch (\beta\mu)+\ch (\beta m)}
+\sgn (eB) \frac{\sh (\beta m)}{\ch (\beta\mu)+\ch (\beta m)}
\right],
\ee
so, extracting parity odd part, one gets for CS at finite temperature
and density
\be
N_{CS}=
\frac{eB}{4\pi}\frac{\sh (\beta m)}{\ch (\beta\mu)+\ch (\beta m)}=
\frac{eB}{4\pi}\th (\beta m)\frac{1}{1+\ch (\beta\mu)/\ch (\beta m)}.
\ee
So, the result coincides with result
for CS term coefficient by Niemi \cite{ni1} obtained for small $\mu$
$\left[ \th \frac{1}{2}\beta(m-\mu)+\th \frac{1}{2}\beta(m+\mu)\right]$.
It is obvious the limit to zero temperature.
The lack of this method is
that it works only for abelian and constant field case.

This result at zero temperature can be obtained by use of
Schwinger proper--time
method. Consider  (2+1) dimensional theory in the abelian case and choose
background field in the form
$$ A^{\mu}=\frac{1}{2}x_{\nu}F^{\nu\mu} ,
\;\;\;\;\;\; F^{\nu\mu}={\rm Const}. $$
To obtain the  CS term in this case, it is necessary to consider
the background current
$$ \langle J^{\mu}\rangle =  \frac{\delta S_{eff}}{\delta A_{\mu}}  $$
rather than the effective action itself. This is because the CS
term formally vanishes for such the choice of
$A^{\mu}$  but its variation with respect to $A^{\mu}$
produces a nonvanishing current.
So, consider
\begin{equation}
\langle J^{\mu}\rangle= -i g \tr\left[\gamma^{\mu}G(x,x^{'})
\right]_{x\rightarrow x^{'}}
\end{equation}
where
\begin{equation}
G(x,x^{'})=\exp\left(-ig\int_{x^{'}}^{x}d\zeta_{\mu}A^{\mu}(\zeta)\right)
\langle x | \hat G | x^{'}\rangle .
\end{equation}
Let's rewrite  Green function (\ref{gre}) in a more appropriate form
\begin{eqnarray}
\hat{G}=(\gamma \tilde{\pi}-m)
\Bigl[ \frac{\theta ((p_{0}+\mu)\sgn (p_{0}))}
{(\gamma\tilde{\pi})^2-m^2 + i \epsilon }+
\frac{\theta (-(p_{0}+\mu)\sgn (p_{0}))}
{(\gamma\tilde{\pi})^2-m^2 - i \epsilon }
\Bigr] .
\end{eqnarray}
Now, we  use the well known integral representation  of denominators
$$\frac{1}{\alpha \pm i0}=\mp i\int_{0}^{\infty}ds\e^{\pm i\alpha s},$$
which corresponds to introducing the ''proper--time'' $s$ into the
calculation
of the Eulier--Hei\-sen\-berg lagrangian by
the Schwinger method \cite{schwin}.
We obtain
\begin{eqnarray}
\hat{G}=(\gamma \tilde{\pi}-m)
\Bigl[ &-&i \int_{0}^{\infty} ds \exp\left( i s \left[
(\gamma\tilde{\pi})^2-m^2 + i \epsilon \right]\right)
\theta ((p_{0}+\mu)\sgn (p_{0}))+\nonumber\\&+&
i \int_{0}^{\infty} ds \exp\left( -i s \left[
(\gamma\tilde{\pi})^2-m^2 - i \epsilon \right]\right)
\theta (-(p_{0}+\mu)\sgn (p_{0}))
\Bigr] .
\end{eqnarray}
For simplicity, we restrict ourselves only to the magnetic field case,
where $A_{0}=0, [\tilde\pi_{0},\tilde\pi_{\mu}]=0 $. Then we easily
can factorize the time dependent part of Green function.
By using the obvious relation
\begin{equation}
\label{yr}
(\gamma\tilde{\pi})^2=(p_0 +\mu)^2 -\vec{\pi}^2 +\frac{1}{2}g
\sigma_{\mu\nu}F^{\mu\nu}
\end{equation}
one gets
\begin{eqnarray}
&\*&G(x,x^{'})|_{x\rightarrow x^{'}}= -i\int\frac{dp_{0}}{2\pi}
\frac{d^2 p}{(2\pi)^2}(\gamma \tilde{\pi}-m)\int_{0}^{\infty} ds
\Biggl[
\e^{is(\tilde{p}_{0}^{2}-m^2)}\e^{-is\vec{\pi}^2}
\e^{isg\sigma F/2}-\nonumber\\
&-&
\theta (-(p_{0}+\mu)\sgn (p_{0}))\left(
\e^{is(\tilde{p}_{0}^{2}-m^2)}\e^{-is\vec{\pi}^2}
\e^{isg\sigma F/2}  +
\e^{-is(\tilde{p}_{0}^{2}-m^2)}\e^{is\vec{\pi}^2}
\e^{-isg\sigma F/2}\right)
\Biggr] .
\end{eqnarray}
Here the first term corresponds to the usual $\mu$--independent case and
there are two additional $\mu$--dependent terms.
In the calculation of the current the following trace arises:
\be
\tr \left[ \gamma^{\mu} (\gamma\tilde{\pi}-m)
\e^{isg\sigma F/2} \right]=
2\pi^{\nu}g^{\nu\mu}\cos ( g|^{*}\! F|s )
+2\frac{\pi^{\nu}F^{\nu\mu}}{|^{*}\! F|}\sin ( g|^{*}\! F|s ) -
2im\frac{^{*}\! F^{\mu}}{|^{*}\! F|}\sin ( g|^{*}\! F|s),\nonumber
\ee
where $\*^{*}\! F^{\mu}=\varepsilon^{\mu\alpha\beta}F_{\alpha\beta}/2$
and $ |^{*}\! F|=\sqrt{B^2-E^2} $.
Since we are interested in  calculation of the
parity odd part (CS term)
it is enough to consider only
terms proportional to the dual strength tensor $\*^{*}\! F^{\mu}$.
On the other hand the term $2\pi^{\nu}g^{\nu\mu}\cos \left( g|^{*}\! F|s\right)$
at $\nu=0$ (see expression for the trace,
we take in mind that here there is only magnetic field)
also gives nonzero contribution to
the current $J^{0}$ \cite{tolpa}
\be
J^{0}_{\rm even}=g\frac{|gB|}{2\pi}\left( {\rm Int}\left[
\frac{\mu^{2}-m^{2}}{2|gB|}
\right]+\frac{1}{2}\right) \theta (|\mu | -|m|).
\ee
This part of current is parity invariant because under parity
$B\rightarrow -B$.
It is clear that this parity even object
does  contribute
neither to the parity anomaly nor to the mass of the
gauge field.
Moreover,  this term has  magnetic field
in the argument's denominator of the cumbersome function --
integer part. So, the parity even term seems to be
''noncovariantizable'', i.e. it can't be converted in covariant
form in effective action.
Since we explore the parity anomalous
topological CS term, we won't consider this parity even term.
So, only the term proportional to the dual strength
tensor $\*^{*}\! F^{\mu}$ gives rise to CS.
The relevant part of the current after
spatial momentum integration  reads
\begin{eqnarray}
J^{\mu}_{CS}=\frac{g^2}{4\pi^2} m \*^{*}\! F^{\mu}
\int_{-\infty}^{+\infty}\!\! dp_{0} \int_{0}^{\infty}\!\! ds
\Biggl[
\e^{is(\tilde{p}_{0}^{2}-m^2)}-
\theta (-\tilde{p}_{0}\sgn (p_{0}))\left(
\e^{is(\tilde{p}_{0}^{2}-m^2)}+
\e^{-is(\tilde{p}_{0}^{2}-m^2)m}\right)\Biggr] .
\end{eqnarray}
Thus, we get besides the usual CS part \cite{redl}, also
the $\mu$--dependent one.
It is easy to calculate it by use of the formula
$$ \int_{0}^{\infty}ds \e^{is(x^2-m^2)}=\pi\left(
\delta (x^2-m^2) +\frac{i}{\pi} {\cal{P}} \frac{1}{x^2-m^2}\right) $$
and we  get eventually
\begin{eqnarray}
\label{fab}
J^{\mu}_{CS}&=&\frac{m}{|m|}\frac{g^2}{4\pi}\*^{*}\! F^{\mu}
\bigl[ 1 - \theta (-(m+\mu)\sgn(m))
-\theta (-(m-\mu)\sgn(m))\bigr]\nonumber\\
&=&\frac{m}{|m|}\theta (m^2 -\mu^2 )\frac{g^2}{4\pi}\*^{*}\! F^{\mu}.
\end{eqnarray}

Let's now discuss the non-abelian case. Then  $A^{\mu}=T_{a} A_{a}^{\mu}$
and
$$\langle J_{a}^{\mu}\rangle= -i g \tr\left[\gamma^{\mu} T_{a}
G(x,x^{'})\right]_{x\rightarrow x^{'}} .$$
It is well--known \cite{redl,red11}
that there exist only two types of the constant background fields.
The first is the  ''abelian'' type
(it is easy to see that the self--interaction
$f^{abc}A_{b}^{\mu}A_{c}^{\mu}$
disappears under that choice of the background field)
\begin{equation}
\label{ab}
A_{a}^{\mu}=\eta_{a}\frac{1}{2}x_{\nu}F^{\nu\mu},
\end{equation}
where $\eta_{a}$ is an arbitrary constant
vector in the color space, $F^{\nu\mu}={\rm Const}$.
The second  is the pure ''non--abelian'' type
\begin{equation}
\label{nab}
A^{\mu}={\rm Const}.
\end{equation}
Here the derivative terms (abelian part) vanish from the strength tensor
and it contains only the self--interaction part
$F^{\mu\nu}_{a}=gf^{abc}A_{b}^{\mu}A_{c}^{\mu}$.
It is clear that to catch  abelian part of the CS term
we should consider the background field (\ref{ab}),
whereas for the non--abelian (derivative noncontaining,
cubic in $A$) part
we have to use the case (\ref{nab}).

Calculations in the ''abelian'' case reduces to the previous analysis,
except the trivial adding  of the color indices in the formula (\ref{fab}):
\begin{eqnarray}
\label{finab}
J^{\mu}_{a}
=\frac{m}{|m|}\theta (m^2 -\mu^2 )
\frac{g^2}{4\pi}\*^{*}F^{\mu}_{a} .
\end{eqnarray}
In the case (\ref{nab})  all calculations are similar. The only
difference is that the origin of term $\sigma_{\mu\nu}F^{\mu\nu}$
in (\ref{yr}) is not the linearity $A$ in $x$ (as in abelian case) but
the pure non--abelian   $A^{\mu}={\rm Const}$. Here  term
$\sigma_{\mu\nu}F^{\mu\nu}$ in (\ref{yr}) becomes quadratic in $A$
and we have
\begin{eqnarray}
\label{finnab}
J^{\mu}_{a}
=\frac{m}{|m|}\theta (m^2 -\mu^2 )
\frac{g^3}{4\pi}\varepsilon^{\mu\alpha\beta}
\tr\left[ T_{a}A^{\alpha}A^{\beta} \right] .
\end{eqnarray}
Combining formulas (\ref{finab}) and (\ref{finnab}) and integrating
over field $A^{\mu}_{a}$  we obtain eventually
\begin{equation}
S^{{\rm CS}}_{eff}=\frac{m}{|m|}\theta (m^2 -\mu^2 ) \pi W[A] ,
\end{equation}
where $W[A]$ is the CS term
$$W[A]=\frac{g^2}{8\pi^2}\int d^{3}x \varepsilon^{\mu\nu\alpha}
\tr \left( F_{\mu\nu}A_{\alpha}-
\frac{2}{3}gA_{\mu}A_{\nu}A_{\alpha}\right) .$$
In conclusion note, that it may seem that covariant notation
used through this section is rather artificial.
However, the covariant notation is useful here because
it helps us to extract Levi-Chivita tensor corresponding
to parity anomalous CS term.

\subsection{Arbitrary gauge field background}
One can see that the procedures we've used above to calculate CS
are noncovariant. Indeed, both of them use the constant
magnetic background.
Here we'll use completely covariant free of any restriction
on gauge field
procedure, which allows us to perform calculations at once
in  nonabelian case.
We'll employ the perturbative expansion.
The zero temperature case within this procedure has been
explored in \cite{my}.

Let's first consider nonabelian 3--dimensional gauge theory.
The only graphs whose P-odd parts  contribute to the
parity anomalous CS term are shown in Fig.1.

\unitlength=1.00mm
\special{em:linewidth 0.4pt}
\linethickness{0.4pt}
\begin{picture}(80.00,50.00)
\put(25.00,35.00){\circle{10.20}}
\put(65.00,35.00){\circle{10.00}}
\put(69.00,32.00){\line(1,0){1.00}}
\put(70.00,32.00){\line(0,-1){1.00}}
\put(70.00,31.00){\line(1,0){1.00}}
\put(71.00,31.00){\line(0,-1){1.00}}
\put(71.00,30.00){\line(1,0){1.00}}
\put(72.00,30.00){\line(0,-1){1.00}}
\put(72.00,29.00){\line(1,0){1.00}}
\put(73.00,29.00){\line(0,-1){1.00}}
\put(73.00,28.00){\line(1,0){1.00}}
\put(74.00,28.00){\line(0,-1){1.00}}
\put(61.00,32.00){\line(-1,0){1.00}}
\put(60.00,32.00){\line(0,-1){1.00}}
\put(60.00,31.00){\line(-1,0){1.00}}
\put(59.00,31.00){\line(0,-1){1.00}}
\put(59.00,30.00){\line(-1,0){1.00}}
\put(58.00,30.00){\line(0,-1){1.00}}
\put(58.00,29.00){\line(-1,0){1.00}}
\put(57.00,29.00){\line(0,-1){1.00}}
\put(57.00,28.00){\line(-1,0){1.00}}
\put(56.00,28.00){\line(0,-1){1.00}}
\put(22.00,20.00){\makebox(0,0)[lb]{(a)}}
\put(63.00,20.00){\makebox(0,0)[lb]{(b)}}
\put(2.00,10.00){\makebox(0,0)[lb]{
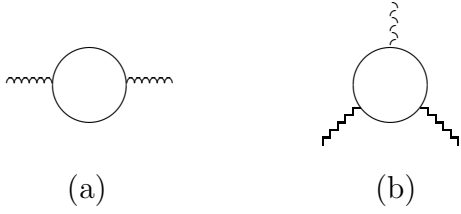
\footnotesize {Fig.1  Graphs whose P-odd parts contribute to
the CS term in nonabelian 3D gauge theory}}}
\put(30.50,35.50){\oval(1.00,1.00)[t]}
\put(31.50,35.50){\oval(1.00,1.00)[t]}
\put(32.50,35.50){\oval(1.00,1.00)[t]}
\put(33.50,35.50){\oval(1.00,1.00)[t]}
\put(34.50,35.50){\oval(1.00,1.00)[t]}
\put(35.50,35.50){\oval(1.00,1.00)[t]}
\put(14.50,35.50){\oval(1.00,1.00)[t]}
\put(15.50,35.50){\oval(1.00,1.00)[t]}
\put(16.50,35.50){\oval(1.00,1.00)[t]}
\put(17.50,35.50){\oval(1.00,1.00)[t]}
\put(18.50,35.50){\oval(1.00,1.00)[t]}
\put(19.50,35.50){\oval(1.00,1.00)[t]}
\put(65.50,40.50){\oval(1.00,1.00)[lt]}
\put(65.50,41.50){\oval(1.00,1.00)[rt]}
\put(65.50,42.50){\oval(1.00,1.00)[lt]}
\put(65.50,43.50){\oval(1.00,1.00)[rt]}
\put(65.50,44.50){\oval(1.00,1.00)[lt]}
\put(65.50,45.50){\oval(1.00,1.00)[rt]}
\end{picture}

So, the part of effective action containing the CS term looks as
\be
\label{eff}
I^{C.S.}_{eff} &=&
\frac{1}{2}\int_{x}  A_{\mu}(x)\int_{p}\e^{-ixp}A_{\nu}(p)
\Pi^{\mu\nu}(p)\nonumber\\ &+&
\frac{1}{3}\int_{x}  A_{\mu}(x)\int_{p,r}\e^{-ix(p+r)}
A_{\nu}(p)A_{\alpha}(r)\Pi^{\mu\nu\alpha}(p,r),
\ee
where polarization operator and vertice have a standard form
\be
\Pi^{\mu\nu}(p) &=&g^2 \int_{k} \tr \left[ \gamma^{\mu}S(p+k;\mu)
\gamma^{\nu}S(k;\mu)\right] \nonumber\\
\Pi^{\mu\nu\alpha}(p,r) &=& g^3
\int_{k}\tr \left[ \gamma^{\mu}S(p+r+k;\mu)
\gamma^{\nu}S(r+k;\mu)\gamma^{\alpha}S(k;\mu)
\right],
\ee
where $S(k;\mu)$ is the Euclidian fermion propagator at finite density
and temperature (\ref{euc}) and  the following notation 
is used $\int_{x}=i\int_{0}^{\beta}dx_{0} \int d\vec{x}$ 
and $\int_{k}=\frac{i}{\beta}\sum_{n=-\infty}^{\infty} \int 
\frac{d\vec{k}}{(2\pi)^{2}}$.  First consider the second order term (Fig.1, 
graph (a)).  It is well-known that the only object giving us the possibility 
to construct $P$ and $T$ odd form in action is Levi-Chivita tensor\footnote{In 
three dimensions it arises as a trace  of three $\gamma$--matrices (Pauli 
matrices)}. Thus, we will drop all terms noncontaining Levi-Chivita tensor.  
Signal for the mass generation (CS term) is
$\Pi^{\mu\nu}(p^{2}=0)\not =0$. So  we  get
\be
\Pi^{\mu\nu}=g^2 \int_{k} ( -i2m e^{\mu\nu\alpha} p_{\alpha} )
\frac{1}{(\tilde{k}^2 +m^2 )^2}.
\ee
After some simple algebra one  obtains
\be
\Pi^{\mu\nu}=-i2mg^2e^{\mu\nu\alpha} p_{\alpha}
\frac{i}{\beta}\sum_{n=-\infty}^{\infty}\int
\frac{d^{2}k}{(2\pi)^2}
\frac{1}{(\tilde{k}^2 +m^2 )^2} =
-i2mg^2e^{\mu\nu\alpha} p_{\alpha}
\frac{i}{\beta}\sum_{n=-\infty}^{\infty}\frac{i}{4\pi}
\frac{1}{\omega_{n}^{2}+m^2},
\ee
where $\omega_{n}=(2n+1)\pi / \beta +i\mu$. Performing
summation we get
\be
\label{6}
\Pi^{\mu\nu}=i\frac{g^2}{4\pi}e^{\mu\nu\alpha} p_{\alpha}
\th (\beta m)\frac{1}{1+\ch (\beta\mu)/\ch (\beta m)}
\ee
It is easily seen that in $\beta\rightarrow\infty$ limit we'll
get zero temperature result \cite{my}
\be
\Pi^{\mu\nu}&=&i\frac{m}{|m|}\frac{g^2}{4\pi}e^{\mu\nu\alpha} p_{\alpha}
\theta (m^2 -\mu^2 ) .
\ee
In the same manner handling the third order contribution (Fig.1b)
one gets
\be
\Pi^{\mu\nu\alpha} &=& -2g^3 i e^{\mu\nu\alpha}
\frac{i}{\beta}\sum_{n=-\infty}^{\infty}\int\frac{d^2 k}{(2\pi)^2}
\frac{m(\tilde{k}^2+m^2)}{ (\tilde{k}^2 +m^2 )^3}
=\nonumber\\&=&-i2mg^3  e^{\mu\nu\alpha}
\frac{i}{\beta}\sum_{n=-\infty}^{\infty}\int\frac{d^2 k}{(2\pi)^2}
\frac{1}{ ( \tilde{k}^2 +m^2 )^2 }
\ee
and  further all calculations are identical to the second order
\be
\label{8}
\Pi^{\mu\nu\alpha}
&=&i\frac{g^3}{4\pi}e^{\mu\nu\alpha}
\th (\beta m)\frac{1}{1+\ch (\beta\mu)/\ch (\beta m)}.
\ee
Substituting (\ref{6}), (\ref{8}) in the effective action
(\ref{eff}) we get eventually
\be
I^{C.S.}_{eff} =\th (\beta m)\frac{1}{1+\ch (\beta\mu)/\ch (\beta m)}
\frac{g^2}{8\pi}
\int d^{3}x e^{\mu\nu\alpha} \tr\left(
A_{\mu}\partial_{\nu}A_{\alpha}-
\frac{2}{3}g A_{\mu}A_{\nu}A_{\alpha}\right).
\ee
Thus, we've got CS term with temperature and density dependent coefficient.

\section{Chern-Simons in arbitrary odd dimension}

Let's now consider 5--dimensional gauge theory.
Here the Levi-Chivita ten\-sor is 5--dimen\-sional $e^{\mu\nu\alpha\beta\gamma}$
and the rele\-vant graphs are shown in Fig.2.

\unitlength=1.00mm
\special{em:linewidth 0.4pt}
\linethickness{0.4pt}
\begin{picture}(120.00,50.00)
\put(25.00,35.00){\circle{10.20}}
\put(65.00,35.00){\circle{10.00}}
\put(105.00,35.00){\circle{10.00}}
\put(25.50,40.50){\oval(1.00,1.00)[lt]}
\put(25.50,41.50){\oval(1.00,1.00)[rt]}
\put(25.50,42.50){\oval(1.00,1.00)[lt]}
\put(25.50,43.50){\oval(1.00,1.00)[rt]}
\put(25.50,44.50){\oval(1.00,1.00)[lt]}
\put(25.50,45.50){\oval(1.00,1.00)[rt]}
\put(105.50,40.50){\oval(1.00,1.00)[lt]}
\put(105.50,41.50){\oval(1.00,1.00)[rt]}
\put(105.50,42.50){\oval(1.00,1.00)[lt]}
\put(105.50,43.50){\oval(1.00,1.00)[rt]}
\put(105.50,44.50){\oval(1.00,1.00)[lt]}
\put(105.50,45.50){\oval(1.00,1.00)[rt]}
\put(23.00,20.00){\makebox(0,0)[lb]{(a)}}
\put(63.00,20.00){\makebox(0,0)[lb]{(b)}}
\put(103.00,20.00){\makebox(0,0)[lb]{(c)}}
\put(3.00,10.00){\makebox(0,0)[lb]{
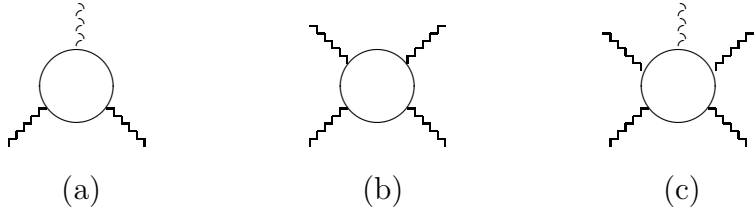
\footnotesize {Fig.2  Graphs whose P-odd parts contribute to the
CS term in nonabelian 5D theory}}}
\put(21.00,32.00){\line(-1,0){1.00}}
\put(20.00,32.00){\line(0,-1){1.00}}
\put(20.00,31.00){\line(-1,0){1.00}}
\put(19.00,31.00){\line(0,-1){1.00}}
\put(19.00,30.00){\line(-1,0){1.00}}
\put(18.00,30.00){\line(0,-1){1.00}}
\put(18.00,29.00){\line(-1,0){1.00}}
\put(17.00,29.00){\line(0,-1){1.00}}
\put(17.00,28.00){\line(-1,0){1.00}}
\put(16.00,28.00){\line(0,-1){1.00}}
\put(29.00,32.00){\line(1,0){1.00}}
\put(30.00,32.00){\line(0,-1){1.00}}
\put(30.00,31.00){\line(1,0){1.00}}
\put(31.00,31.00){\line(0,-1){1.00}}
\put(31.00,30.00){\line(1,0){1.00}}
\put(32.00,30.00){\line(0,-1){1.00}}
\put(32.00,29.00){\line(1,0){1.00}}
\put(33.00,29.00){\line(0,-1){1.00}}
\put(33.00,28.00){\line(1,0){1.00}}
\put(34.00,28.00){\line(0,-1){1.00}}
\put(61.00,32.00){\line(-1,0){1.00}}
\put(60.00,32.00){\line(0,-1){1.00}}
\put(60.00,31.00){\line(-1,0){1.00}}
\put(59.00,31.00){\line(0,-1){1.00}}
\put(59.00,30.00){\line(-1,0){1.00}}
\put(58.00,30.00){\line(0,-1){1.00}}
\put(58.00,29.00){\line(-1,0){1.00}}
\put(57.00,29.00){\line(0,-1){1.00}}
\put(57.00,28.00){\line(-1,0){1.00}}
\put(56.00,28.00){\line(0,-1){1.00}}
\put(101.00,32.00){\line(-1,0){1.00}}
\put(100.00,32.00){\line(0,-1){1.00}}
\put(100.00,31.00){\line(-1,0){1.00}}
\put(99.00,31.00){\line(0,-1){1.00}}
\put(99.00,30.00){\line(-1,0){1.00}}
\put(98.00,30.00){\line(0,-1){1.00}}
\put(98.00,29.00){\line(-1,0){1.00}}
\put(97.00,29.00){\line(0,-1){1.00}}
\put(97.00,28.00){\line(-1,0){1.00}}
\put(96.00,28.00){\line(0,-1){1.00}}
\put(74.00,43.00){\line(-1,0){1.00}}
\put(73.00,43.00){\line(0,-1){1.00}}
\put(73.00,42.00){\line(-1,0){1.00}}
\put(72.00,42.00){\line(0,-1){1.00}}
\put(72.00,41.00){\line(-1,0){1.00}}
\put(71.00,41.00){\line(0,-1){1.00}}
\put(71.00,40.00){\line(-1,0){1.00}}
\put(70.00,40.00){\line(0,-1){1.00}}
\put(70.00,39.00){\line(-1,0){1.00}}
\put(69.00,39.00){\line(0,-1){1.00}}
\put(115.00,42.00){\line(-1,0){1.00}}
\put(114.00,42.00){\line(0,-1){1.00}}
\put(114.00,41.00){\line(-1,0){1.00}}
\put(113.00,41.00){\line(0,-1){1.00}}
\put(113.00,40.00){\line(-1,0){1.00}}
\put(112.00,40.00){\line(0,-1){1.00}}
\put(112.00,39.00){\line(-1,0){1.00}}
\put(111.00,39.00){\line(0,-1){1.00}}
\put(111.00,38.00){\line(-1,0){1.00}}
\put(110.00,38.00){\line(0,-1){1.00}}
\put(56.00,43.00){\line(1,0){1.00}}
\put(57.00,43.00){\line(0,-1){1.00}}
\put(57.00,42.00){\line(1,0){1.00}}
\put(58.00,42.00){\line(0,-1){1.00}}
\put(58.00,41.00){\line(1,0){1.00}}
\put(59.00,41.00){\line(0,-1){1.00}}
\put(59.00,40.00){\line(1,0){1.00}}
\put(60.00,40.00){\line(0,-1){1.00}}
\put(60.00,39.00){\line(1,0){1.00}}
\put(61.00,39.00){\line(0,-1){1.00}}
\put(95.00,42.00){\line(1,0){1.00}}
\put(96.00,42.00){\line(0,-1){1.00}}
\put(96.00,41.00){\line(1,0){1.00}}
\put(97.00,41.00){\line(0,-1){1.00}}
\put(97.00,40.00){\line(1,0){1.00}}
\put(98.00,40.00){\line(0,-1){1.00}}
\put(98.00,39.00){\line(1,0){1.00}}
\put(99.00,39.00){\line(0,-1){1.00}}
\put(99.00,38.00){\line(1,0){1.00}}
\put(100.00,38.00){\line(0,-1){1.00}}
\put(69.00,32.00){\line(1,0){1.00}}
\put(70.00,32.00){\line(0,-1){1.00}}
\put(70.00,31.00){\line(1,0){1.00}}
\put(71.00,31.00){\line(0,-1){1.00}}
\put(71.00,30.00){\line(1,0){1.00}}
\put(72.00,30.00){\line(0,-1){1.00}}
\put(72.00,29.00){\line(1,0){1.00}}
\put(73.00,29.00){\line(0,-1){1.00}}
\put(73.00,28.00){\line(1,0){1.00}}
\put(74.00,28.00){\line(0,-1){1.00}}
\put(109.00,32.00){\line(1,0){1.00}}
\put(110.00,32.00){\line(0,-1){1.00}}
\put(110.00,31.00){\line(1,0){1.00}}
\put(111.00,31.00){\line(0,-1){1.00}}
\put(111.00,30.00){\line(1,0){1.00}}
\put(112.00,30.00){\line(0,-1){1.00}}
\put(112.00,29.00){\line(1,0){1.00}}
\put(113.00,29.00){\line(0,-1){1.00}}
\put(113.00,28.00){\line(1,0){1.00}}
\put(114.00,28.00){\line(0,-1){1.00}}
\end{picture}

The part of effective action containing CS term reads
\be
\label{ef}
I^{C.S.}_{eff} &=&
\frac{1}{3}\int_{x}  A_{\mu}(x)\int_{p,r}\e^{-ix(p+r)}
A_{\nu}(p)A_{\alpha}(r)\Pi^{\mu\nu\alpha}(p,r)\nonumber\\
&+&
\frac{1}{4}\int_{x}  A_{\mu}(x)\int_{p,r,s}\e^{-ix(p+r+s)}
A_{\nu}(p)A_{\alpha}(r)A_{\beta}(s)\Pi^{\mu\nu\alpha\beta}(p,r,s)\nonumber\\
&+&
\frac{1}{5}\int_{x}  A_{\mu}(x)\int_{p,r,s,q}\e^{-ix(p+r+s+q)}
A_{\nu}(p)A_{\alpha}(r)A_{\beta}(s)A_{\gamma}(s)\Pi^{\mu\nu\alpha\beta\gamma}
(p,r,s,q)
\ee
All calculations are similar to 3--dimensional case.
First consider third order contribution (Fig.2a)
\be
\Pi^{\mu\nu\alpha}(p,r) = g^3
\int_{k}\tr \left[ \gamma^{\mu}S(p+r+k;\mu)
\gamma^{\nu}S(r+k;\mu)\gamma^{\alpha}S(k;\mu)
\right].
\ee
Taking into account that trace of five $\gamma$-matrices in
5--dimensions is
$$
\tr\left[\gamma^{\mu}\gamma^{\nu}\gamma^{\alpha}\gamma^{\beta}\gamma^{\rho}
\right] = 4ie^{\mu\nu\alpha\beta\rho}, $$
we  extract the parity odd part of the vertice
\be
\Pi^{\mu\nu\alpha}=g^3 \frac{i}{\beta}\sum_{n=-\infty}^{\infty}
\int\frac{d^4 k}{(2\pi)^4}
( i4m e^{\mu\nu\alpha\beta\sigma} p_{\beta}r_{\sigma} )
\frac{1}{(\tilde{k}^2 +m^2 )^3},
\ee
or in more transparent way
\be
\Pi^{\mu\nu\alpha}&=& i4mg^3 e^{\mu\nu\alpha\beta\sigma}
p_{\alpha}r_{\sigma} \frac{i}{\beta} \sum_{n=-\infty}^{+\infty}
\int\frac{d^{4}k}{(2\pi)^4}
\frac{1}{(\omega_{n}^2 +\vec{k}^2 +m^2 )^3}=\nonumber\\&=&
i4mg^3 e^{\mu\nu\alpha\beta\sigma}
p_{\alpha}r_{\sigma} \frac{i}{\beta} \sum_{n=-\infty}^{+\infty}
\frac{-i}{64\pi^2} \frac{1}{\omega_{n}^2+m^2}.
\ee
Performing summation one comes to
\be
\label{p1}
\Pi^{\mu\nu\alpha}
=i\th (\beta m)\frac{1}{1+\ch (\beta\mu)/\ch (\beta m)}
\frac{g^3}{16\pi^2}e^{\mu\nu\alpha\beta\sigma}
p_{\alpha}r_{\sigma}.
\ee
In the same way operating graphs (b) and (c) (Fig.2) one will obtain
\be
\Pi^{\mu\nu\alpha\beta}
=i\th (\beta m)\frac{1}{1+\ch (\beta\mu)/\ch (\beta m)}
\frac{g^4}{8\pi^2}e^{\mu\nu\alpha\beta\sigma}s_{\sigma}
\ee
and
\be
\label{p2}
\Pi^{\mu\nu\alpha\beta\gamma}
=i\th (\beta m)\frac{1}{1+\ch (\beta\mu)/\ch (\beta m)}
\frac{g^5}{16\pi^2}e^{\mu\nu\alpha\beta\sigma}.
\ee
Substituting (\ref{p1}) --- (\ref{p2}) in the effective
action (\ref{ef}) we get the
final result for  CS in 5--dimensional theory
\be
I^{C.S.}_{eff} &=&\th (\beta m)\frac{1}{1+\ch (\beta\mu)/\ch (\beta m)}
\frac{g^3}{48\pi^2}
\int_{x} e^{\mu\nu\alpha\beta\gamma} \nonumber\\
&\times&\tr\left(
A_{\mu}\partial_{\nu}A_{\alpha}\partial_{\beta}A_{\gamma}+
\frac{3}{2}g A_{\mu}A_{\nu}A_{\alpha}\partial_{\beta}A_{\gamma}+
\frac{3}{5}g^{2} A_{\mu}A_{\nu}A_{\alpha}A_{\beta}A_{\gamma}
\right).
\ee

It is remarkable that all parity odd contributions are finite
as in 3--dimensional so as in 5--dimensional cases.
Thus, all values in the effective action are renormalized in
a standard way, i.e. the renormalizations are
determined by conventional (parity even) parts of vertices.

From the above direct calculations  it is clearly seen
that the chemical potential and temperature dependent coefficient is the
same for all parity odd parts of diagrams and doesn't depend on
space dimension. So, the  influence
of finite density and temperature on CS term generation is the same in
any odd dimension:
\begin{equation}
\label{kon}
I^{{\rm C.S}}_{eff}=\th (\beta m)\frac{1}{1+\ch (\beta\mu)/\ch (\beta m)}
 \pi W[A] \stackrel{\beta\rightarrow\infty}{\longrightarrow}
\frac{m}{|m|}\theta (m^2 -\mu^2 )\pi W[A],
\end{equation}
where $W[A]$ is the CS secondary characteristic class
in any odd dimension.
Since  only the lowest orders
of perturbative series contribute to CS term at finite density and temperature
(the same situation
is well-known at zero density), the result obtained by using
formally perturbative technique appears to be nonperturbative.
Thus, the $\mu$ and $T$ --dependent CS term coefficient
reveals the amazing property of universality.
Namely, it does depend on
neither dimension of the theory nor abelian or nonabelian gauge
theory is studied.

The arbitrariness of $\mu$ gives us the possibility
to see CS coefficient behavior at any masses.
It is very interesting that  $\mu^2 = m^2$ is the
crucial point for CS at zero temperature.
Indeed, it is clearly seen from (\ref{kon}) that when $\mu^2 < m^2$
$\mu$--influence disappears and we get the usual CS term
$I^{{\rm C.S}}_{eff}= \pi W[A].$
On the other hand when $\mu^2 > m^2$
the situation is absolutely different.
One can see that here the CS term
disappears because of non--zero density of background fermions.
We'd like to emphasize the
important massless case $m=0$ considered in many a papers,
see for example \cite{ni1,redl,jac}.
Here even negligible density or temperature,
which always take place in any
physical processes, leads to vanishing of the parity anomaly.
Let us stress again that we nowhere have used
any restrictions on $\mu$.
Thus we not only confirm result
in \cite{ni1} for CS in $QED_{3}$ at small density,
but also expand it
on arbitrary $\mu$, nonabelian case and arbitrary odd dimension.

\section{Trace identity}
Here, we'll consider trace identity at finite temperature
and density. First of all, by using well-known trace identity
at finite temperature \cite{ni,ni1},
we'll present the simple resons that chiral anomaly
doesn't depend on temperature
in any even dimension. Indeed, at finite temperature and zero density
trace identity still holds  and one has \cite{ni,ni1}
\be
\label{in}
\langle N\rangle_{\beta}=-\frac{1}{2\beta}\sum_{-\infty}^{+\infty}
\frac{m}{m^2 +\omega^{2}_{n}}\left(\int dx ({\rm anomaly}) +
\int dx \partial_{i}\tr\langle x |i\Gamma_{i}\Gamma^{c}
\frac{1}{H_{0}+i\sqrt{m^2 + \omega^{2}_{n}}}\rangle\right).
\ee
The second term at left hand side is a surface term, which doesn't
contribute to topological part of the trace identity \cite{ni,ni1}.
Thus, for topological part, we are interested in, trace
identity takes the form
\be
\label{in1}
\langle N\rangle_{\beta}^{\rm topological}
=-\frac{1}{2\beta}\sum_{-\infty}^{+\infty}
\frac{m}{m^2 +\omega^{2}_{n}}\left(\int dx ({\rm anomaly})
\right).
\ee
The result for left hand side of  eq.(\ref{in1}) we know
in arbitrary odd dimension.  Really, substituting (\ref{kon}) in
\be
\label{dl}
\langle N\rangle_{\beta}^{\rm CS}=
\langle N\rangle_{\beta}^{\rm topological}
=\frac{\delta I^{{\rm C.S}}_{eff}}{g\delta A_{0}},
\ee
and taking into account that
\be
\frac{1}{2\beta} \sum_{n=-\infty}^{+\infty}
\frac{m}{\omega_{n}^2+m^2} =\frac{1}{4}\frac{\sh (\beta m)}
{1+\ch (\beta m)},
\ee
one can see that the only possibility to reconcile left and right
sides of eq.(\ref{in1}) is to put temperature independence of
anomaly. Thus, we've got that axial anomaly doesn't depend on
temperature in any even-dimensional theory.

Further,  we can generalize trace identity for topological part
on arbitrary finite density.
Really, from (\ref{kon}) and (\ref{dl}) we get
\be
\label{odd}
\langle N\rangle_{\beta , \mu}^{\rm CS}= -
\frac{1}{4}\th (\beta m)\frac{1}{1+\ch (\beta\mu)/\ch (\beta m)}
\int dx \left( {\rm anomaly} \right),
\ee
where $\langle N\rangle_{\beta , \mu}^{\rm CS}$ --
odd part of fermion number in $D$-dimensional theory
at finite density and temperature,  $\left( {\rm anomaly} \right)$ --
axial anomaly in $(D-1)$-dimensional theory.
On the other hand, as we have seen above, the anomaly doesn't
depend on $\mu$  in 2 and 4 dimensions (and doesn't
depend on $T$ in any even-dimensional theory).
Our comprehension of the problem allows us to
generalize this on arbitrary even dimension.
Indeed, anomaly is  the result of  ultraviolet regularization,
while $\mu$ (and $T$) don't effect on ultraviolet
behavior of a theory. Taking in mind (\ref{odd})  and
that at finite density
\be
\frac{1}{2\beta} \sum_{n=-\infty}^{+\infty}
\frac{m}{\omega_{n}^2+m^2} =\frac{1}{4}
\th (\beta m)\frac{1}{1+\ch (\beta\mu)/\ch (\beta m)}
\ee
we can identify
$\langle N\rangle_{\beta , \mu}^{\rm topological}$ and
$\langle N\rangle_{\beta , \mu}^{\rm CS}$.
So, we get generalized on finite density trace identity
for topological part of fermion number
\be
\label{g}
\langle N\rangle_{\beta ,\mu}^{\rm CS}=
\langle N\rangle_{\beta , \mu}^{\rm topological}=
-\frac{1}{2\beta}\sum_{-\infty}^{+\infty}
\frac{m}{m^2 +\omega^{2}_{n}}\left(\int dx ({\rm anomaly})\right).
\ee
The physical underground of formula (\ref{g}) could be more clear
understood if we remember calculations we've performed in sect.3.1
by use of summation over Landau levels.
Really, we've seen that only zero modes contribute to $P$-odd part
in contrast to $P$-even part which is contributed by all modes.
Therefore, index theorem and trace identities hold only
for parity odd (topological) part of fermion number at finite density.

Thus, eq.(\ref{g}) connects CS term and chiral anomaly
in arbitrary dimensional
theory at finite density and temperature.

\section{Conclusions}
The   finite temperature and density influence
on CS term generation is obtained in any odd dimensional theory as for
abelian, so as for nonabelian cases.
It is of interest that  $\mu^2 = m^2$ is the
crucial point for CS at zero temperature.
Indeed, it is clearly seen from (\ref{kon}) that when $\mu^2 < m^2$
$\mu$--influence disappears and we get the usual CS term
$I^{{\rm C.S}}_{eff}= \pi W[A].$
On the other hand when $\mu^2 > m^2$
the CS term disappears because of non--zero
density of background fermions.

The $\mu$ and $T$--dependent CS term coefficient reveals
the amasing property of universality. Namely, it does depend on neither
dimension of the theory nor abelian or nonabelian
gauge theory is studied.
It must be stressed that
at $m=0$  even negligible density or temperature,
which always take place in any
physical processes, leads to vanishing of the parity anomaly.

The  medium effects such as finite density and temperature influence
on chiral anomaly have been studied.
The simple and general arguments that chiral anomaly is
independent of temperature have been presented.
It is shown that  even if we introduce
conservation of chiral charge as the constraint, the chiral
anomaly  isn't effected.
By using the fact that chiral anomaly doesn't depend on
temperature and density we explore the CS number
appearance of CS number in even-dimensional theories
under two type of constraints. These are
charge conservation with Lagrange multiplier $\mu$
(conventional chemical potential) and chiral charge
conservation with Lagrange multiplier $\kappa$,
what corresponds to conservation of the left(right)-handed
fermions asymmetry in the background.

On the other hand, the chiral anomaly independence of density
and temperature together with our direct calculations of CS
coefficient permit us the simple generalization of trace identity
on finite density case.
Thus, the connection between CS term
and chiral anomaly at finite density and temperature
is obtained in arbitrary dimensional theory.

\end{document}